\documentclass[twocolumn,english,pra]{revtex4}
\usepackage[T1]{fontenc}
\usepackage[latin9]{inputenc}
\usepackage{amsmath}
\usepackage{graphicx}
\usepackage{amssymb}
\usepackage{esint}

\makeatletter
\@ifundefined{textcolor}{}
{%
 \definecolor{BLACK}{gray}{0}
 \definecolor{WHITE}{gray}{1}
 \definecolor{RED}{rgb}{1,0,0}
 \definecolor{GREEN}{rgb}{0,1,0}
 \definecolor{BLUE}{rgb}{0,0,1}
 \definecolor{CYAN}{cmyk}{1,0,0,0}
 \definecolor{MAGENTA}{cmyk}{0,1,0,0}
 \definecolor{YELLOW}{cmyk}{0,0,1,0}
 }

\makeatother

\makeatother

\usepackage{babel}

\makeatother

\usepackage{babel}

\begin{document}

\title{Three attractively interacting fermions in a harmonic trap: \\
Exact solution, ferromagnetism, and high-temperature thermodynamics}

\author{Xia-Ji Liu$^{1}$}

\email{xiajiliu@swin.edu.au}

\affiliation{$^{1}$ARC Centre of Excellence for Quantum-Atom Optics, Centre for
Atom Optics and Ultrafast Spectroscopy, Swinburne University of Technology,
Melbourne 3122, Australia}

\author{Hui Hu$^{1}$}

\email{hhu@swin.edu.au}

\affiliation{$^{1}$ARC Centre of Excellence for Quantum-Atom Optics, Centre for
Atom Optics and Ultrafast Spectroscopy, Swinburne University of Technology,
Melbourne 3122, Australia}

\author{Peter D. Drummond$^{1}$}

\email{pdrummond@swin.edu.au}

\affiliation{$^{1}$ARC Centre of Excellence for Quantum-Atom Optics, Centre for
Atom Optics and Ultrafast Spectroscopy, Swinburne University of Technology,
Melbourne 3122, Australia}

\date{\today{}}
\begin{abstract}
Three fermions with strongly repulsive interactions in a spherical
harmonic trap, constitute the simplest nontrivial system that can
exhibit the onset of itinerant ferromagnetism. Here, we present exact
solutions for three trapped, attractively interacting fermions near
a Feshbach resonance. We analyze energy levels on the upper branch
of the resonance where the atomic interaction is effectively repulsive.
When the \emph{s}-wave scattering length $a$ is sufficiently positive,
three fully polarized fermions are energetically stable against a
single spin-flip, indicating the possibility of itinerant ferromagnetism,
as inferred in the recent experiment. We also investigate the high-temperature
thermodynamics of a strongly repulsive or attractive Fermi gas using
a quantum virial expansion. The second and third virial coefficients
are calculated. The resulting equations of state can be tested in
future quantitative experimental measurements at high temperatures
and can provide a useful benchmark for quantum Monte Carlo simulations. 
\end{abstract}


\maketitle

\section{Introduction}

Few-particle systems have become increasingly crucial to the physics
of strongly interacting ultracold quantum gases \cite{braaten,bloch,giorgini}.
Because of large interaction parameters, conventional perturbation
theory approaches to quantum gases such as mean-field theory simply
break down \cite{bloch,giorgini,unitaritycmp}. A small ensemble of
a few fermions and/or bosons, which is either exactly solvable or
numerically tractable, is more amenable to \emph{nonperturbative}
quantal calculations. Although challenging experimentally, such ensembles
benefit from the same unprecedented controllability and tunability
as in a mesoscopic system containing a hundred thousand particles.
The atomic species, the quantum statistics, the \textit{s}-wave and
higher partial wave interactions \cite{chin}, and the external trapping
environment can all be controlled experimentally. The study of few-particle
systems can therefore give valuable insights into the more complicated
mesoscopic many-body physics of a strongly interacting quantum gas.
In addition to qualitative insights, these solutions have already
proved invaluable in developing high-temperature quantum virial or
cluster expansions for larger systems \cite{liu2009}, which have
been recently verified experimentally \cite{ensexpt}.

The purpose of this paper is to add a further milestone in this direction.
By exactly solving the eigenfunctions of three attractively interacting
fermions in a spherical harmonic trap, we aim to give a few-body perspective
of itinerant ferromagnetism in an \emph{effectively }repulsive Fermi
gas, which was observed as a transient phenomenon in a recent measurement
at MIT \cite{mitexpt}. This is possible because the quantum three-body
problem with \textit{s}-wave interactions is exactly soluble in three
dimensions. It is interesting to recall that the corresponding classical
three-body problem is notoriously insoluble. The reason for this unexpectedly
docile quantum behaviour is that the \textit{s}-wave interaction Hamiltonian
applicable to ultra-cold Bose and Fermi gases is essentially just
a boundary condition on an otherwise non-interacting quantum gas.
Thus, we have an unusual situation where quantum mechanics actually
simplifies an intractable classical problem.

For Bose-Einstein condensates (BEC) in the strongly interacting regime,
three trapped bosonic atoms with large \textit{s}-wave scattering
length were already investigated theoretically as a minimum prototype
\cite{blume2002} of this few-body physics. To understand the fascinating
crossover from a BEC to a Bardeen-Cooper-Schrieffer (BCS) superfluid,
two spin-up and two spin-down fermions in a trap were also simulated
numerically, constituting the simplest model of the BEC-BCS crossover
problem \cite{stecher2007,blume2009}. Moreover, knowledge of few-particle
processes such as three-body recombination is primarily responsible
for controlling the loss rate or lifetime of ultracold atomic gases,
which, in many cases, imposes severe limitations on experiments. Important
examples in this context include the confirmation of stability of
dimers in the BEC-BCS crossover \cite{petrov} and the discovery of
the celebrated Efimov state (i.e., a bound state of three resonantly
interacting bosons) as well as the related universal four-body bound
state \cite{physics}.

Whether an itinerant Fermi gas with repulsive interactions exhibits
ferromagnetism is a long-standing problem in condensed matter physics
\cite{stoner1938}. It has recently attracted increasing attention
in the cold-atom community \cite{macdonald2005,zhang2008,leblanc2009,conduit2009a,conduit2009b,zhai2009,cui2010,spilati,chang,dong2010}.
The answer depends on a competition between the repulsive interaction
energy and the cost of kinetic energy arising from Pauli exclusion.
A strong repulsive interaction can induce polarization or ferromagnetism,
since fermions with the same spin orientation are protected from local
interactions by the exclusion principle. This, however, increases
the Fermi energy, as all fermions must now occupy the same band. The
difficulty in finding the transition point is that quantum correlations
change the interaction energy in a way that is difficult to calculate
in general.

As early as the 1930's, Stoner showed with a simple model using mean-field
theory that ferromagnetism in a homogeneous Fermi gas will always
take place \cite{stoner1938}. This model, however, gives the unphysical
result that the interaction energy within the mean-field approximation
scales linearly with the $s$-wave scattering length $a$ and therefore
could be infinitely large. The predicted critical interaction strength
at zero temperature, $(k_{F}a)_{c}=\pi/2$, where $k_{F}$ is the
Fermi wave-vector is also too large in the mean-field picture. An
improved prediction from second-order perturbation theory, $(k_{F}a)_{c}\simeq1.054$,
suffers from similar doubts about its validity \cite{macdonald2005,conduit2009a}.
Most recently, three independent {\em ab-initio} quantum Monte
Carlo simulations conclusively reported a zero-temperature ferromagnetic
phase transition at $(k_{F}a)_{c}\simeq0.8-0.9$ \cite{conduit2009b,spilati,chang}. 

Several important issues are still open, including the nature of transition
at finite temperatures. The unitarity limited interaction energy at
infinitely large scattering length ($a\rightarrow\infty$) is also
to be determined.

\begin{figure}
\begin{centering}
\includegraphics[clip,width=0.45\textwidth]{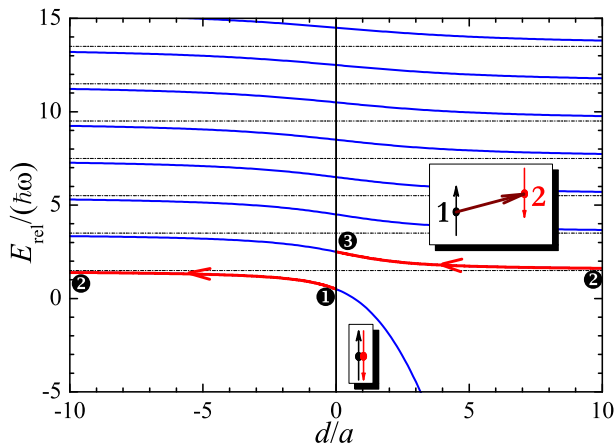} 
\par\end{centering}

\caption{(Color online) Energy spectrum of the relative motion of a trapped
two-fermion system near a Feshbach resonance (i.e, $d/a=0$, where
$d$ is the characteristic harmonic oscillator length). For a positive
scattering length $a>0$ in the right part of the figure, the ground
state is a molecule with size $a$, whose energy diverges as $E_{rel}\simeq-\hbar^{2}/(ma^{2})$.
The excited states or the upper branch of the resonance may be viewed
as the Hilbert space of a {}``repulsive'' Fermi gas with the same
scattering length $a$. In this two-body picture, the level from the
point 2 to 3 is the ground state energy level of the repulsive two-fermion
sub-space, whose energy initially increases linearly with increasing
$a$ from $1.5\hbar\omega$ at the point 2 and finally saturates towards
$2.5\hbar\omega$ at the resonance point 3. For comparison, we illustrate
as well the ground state energy level in the case of a negative scattering
length and show how the energy increases with increased scattering
length from point 1 to 2.}

\label{fig1} 
\end{figure}

The exciting experiment at MIT of $^{6}$Li atoms realized in some
sense the textbook Stoner model \cite{mitexpt}. A crucial aspect
in this realization is that the interatomic interactions are very
different from the conventional model of hard-sphere interactions
\cite{macdonald2005,spilati,chang}. In the experiment, the atoms
are on the {\em upper branch} of a Feshbach resonance with a positive
scattering length $a>0$ and a negligible effective interaction range.
The properties of the atoms are therefore universal, independent of
the details of the interactions \cite{ho2004a,natphys}. This universality,
however, comes at a price: the underlying two-body interaction is
always attractive, so that the ground state is a gas of molecules
of size $a$. The experiment thus suffers from considerable atom loss
and has to be carried out under nonequilibirum conditions. This is
clearly explained in Fig. 1, which shows the relative energy spectrum
of a pair of fermions in a harmonic trap with frequency $\omega$
across a Feshbach resonance \cite{busch}. For $a>0$, the whole spectrum
consists of two distinct parts, the lowest ground-state branch diverging
as $E_{rel}\simeq-\hbar^{2}/(ma^{2})$ and the regular upper branch
having a finite energy. In the context of this {\em two-body} picture,
an $s$-wave {}``repulsive'' Fermi gas is realized, provided that
there are no pairs of fermions occupying the ground state branch of
molecules. However, as far as the many-particle aspect is concerned,
it is not clear to what extent this two-body picture of a {}``repulsive''
Fermi gas will persist. In other words, can we prove rigorously that
the whole Hilbert space of an attractive many-fermion system with
a positive scattering length consists exactly of a sub-Hilbert space
of a repulsive Fermi gas with the same scattering length, together
with an orthogonalized subspace of molecules?

This paper addresses the problem of itinerant ferromagnetism in an
attractive Fermi gas using a few-particle perspective, by examining
the exact solutions for the energy spectrum of three trapped, attractively
interacting fermions in their upper branch of the Feshbach resonance.
Our main results may be summarized as follows.
\begin{itemize}
\item First, we present an elegant and physically transparent way to exactly
solve the Hamiltonian of three interacting fermions in a harmonic
trap. The method may easily be generalized to treat other systems
with different types of atomic species, geometries, and interactions.
\item Secondly, we observe clearly from the whole relative energy spectrum
of three attractive fermions (see Fig. 3) that, there are indeed two
branches of the spectrum on the side of positive scattering length.
As the scattering length goes to an infinitely small positive value,
the lower branch diverges in energy to $-\infty$, while another upper
branch always converges to the non-interacting limit. The latter may
be interpreted as the energy spectrum of three {}``repulsively''
interacting fermions. However, close to the Feshbach resonance, there
are many nontrivial avoided crossings between two types of spectrum,
making it difficult to unambiguously identify a repulsive Fermi system.
These avoided crossings are expected in more general cases, and lead
to nontrivial consequences in a time-dependent field-sweep experiment
passing from the weakly interacting regime at $a=0^{+}$ to the unitarity
limit at $a=+\infty$.
\item Thirdly, we show exactly that near the Feshbach resonance, three {}``repulsively''
interacting fermions in a spherical harmonic trap, say, in a two spin-up
and one spin-down configuration, are higher in total energy than three
fully polarized fermions (see the ground state energy in Fig. 3b).
Thus, there must be a ferromagnetic transition occurring at a certain
critical scattering length. Note that ferromagnetism cannot be obtained
in a two-fermion system. As shown in Fig. 1, even at resonance the
total ground state energy of a repulsively interacting pair, $E_{pair}=4\hbar\omega$,
which is the sum of the relative ground state energy $E_{rel}=2.5\hbar\omega$
and the zero-point energy of center-of-mass motion $E_{cm}=1.5\hbar\omega$,
cannot be larger than the total ground state energy of two fully polarized
fermions, i.e., $E_{\uparrow\uparrow}=4\hbar\omega$.
\item Last but most importantly, we obtain the high temperature equations
of state of strongly interacting Fermi gases (see Figs. 4, 5, 6, and
7), within a quantum virial expansion theory, which was developed
recently by the present authors \cite{liu2009}. The second and third
virial (expansion) coefficients of both attractive and repulsive Fermi
gases can be calculated, using the full energy spectrum of three interacting
fermions. In the unitarity limit, we find that in the high temperature
regime where our quantum virial expansion is reliable, the itinerant
ferromagnetism disappears.
\end{itemize}
The paper is organized as follows. In the next section, we outline
the theoretical model for a few fermions with $s$-wave interactions
in a spherical harmonic trap. In Sec. III, we explain how to construct
the exact wavefunctions for three interacting fermions and discuss
in detail the whole energy spectrum. In Sec. IV we develop a quantum
virial expansion for thermodynamics and calculate the second and third
virial coefficients, based on the full energy spectrum of two-fermion
and three-fermion systems, respectively. The high temperature equations
of state of strongly interacting Fermi gases are then calculated and
discussed in Sec. V. Finally, Sec. VI is devoted to conclusions and
some final remarks. The appendix shows the numerical details of the
exact solutions.

\section{Models}

Consider a few fermions in a three-dimensional (3D) {\em isotropic}
harmonic trap $V({\bf x})=m\omega^{2}x^{2}/2$ with the same mass
$m$ and trapping frequency $\omega$, occupying two different hyperfine
states or two spin states. The fermions with {\em unlike} spins
attract each other via a short-range $s$-wave contact interaction.
It is convenient to use the Bethe-Peierls boundary condition to replace
the $s$-wave pseudopotential. That is, when any particles $i$ and
$j$ with unlike spins close to each other, $r_{ij}=\left|{\bf x}_{i}-{\bf x}_{j}\right|\rightarrow0$,
the few-particle wave function $\psi\left({\bf x}_{1},{\bf x}_{2},...,{\bf x}_{N}\right)$
with proper symmetry should satisfy \cite{wernerprl,wernerpra,footnote},
\begin{equation}
\psi={\cal A}_{ij}({\bf X}_{ij}=\frac{{\bf x}_{i}+{\bf x}_{j}}{2},\{{\bf x}_{k\neq i,j}\})\left(\frac{1}{r_{ij}}-\frac{1}{a}\right),\end{equation}
 where ${\cal A}_{ij}({\bf X}_{ij},\{{\bf x}_{k\neq i,j}\})$ is a
function independent of $r_{ij}$, and $a$ is $s$-wave scattering
length. This boundary condition can be equivalently written as, \begin{equation}
\lim_{r_{ij}\rightarrow0}\frac{\partial\left(r_{ij}\psi\right)}{\partial r_{ij}}=-\frac{r_{ij}\psi}{a}.\label{BP}\end{equation}
 Otherwise, the wave function $\psi$ obeys a non-interacting Schrödinger
equation, \begin{equation}
\sum_{i=1}^{N}\left[-\frac{\hbar^{2}}{2m}{\bf \nabla}_{{\bf x}_{i}}^{2}+\frac{1}{2}m\omega^{2}x_{i}^{2}\right]\psi=E\psi.\end{equation}
 We now describe how to solve all the wave functions with energy level
$E$ for a two- or three-fermion system.

\section{Method}

In a harmonic trap, it is useful to separate the center-of-mass motion
and relative motion. We thus define the following center-of-mass coordinate
${\bf R}$ and relative coordinates ${\bf r}_{i}$ ($i\geq2$) for
$N$ fermions in a harmonic trap \cite{wernerpra,footnote}, \begin{equation}
{\bf R}=\left({\bf x}_{1}+\cdots+{\bf x}_{N}\right)/N,\end{equation}
 and \begin{equation}
{\bf r}_{i}=\sqrt{\frac{i-1}{i}}\left({\bf x}_{i}-\frac{1}{i-1}\sum_{k=1}^{i-1}{\bf x}_{k}\right),\end{equation}
 respectively. In this Jacobi coordinate, the Hamiltonian of the non-interacting
Schrödinger equation takes the form ${\cal H}_{0}={\cal H}_{cm}+{\cal H}_{rel}$,
where, \begin{equation}
{\cal H}_{cm}=-\frac{\hbar^{2}}{2M}{\bf \nabla}_{{\bf R}}^{2}+\frac{1}{2}M\omega^{2}R^{2},\end{equation}
 and \begin{equation}
{\cal H}_{rel}=\sum_{i=2}^{N}\left[-\frac{\hbar^{2}}{2m}{\bf \nabla}_{{\bf r}_{i}}^{2}+\frac{1}{2}m\omega^{2}r_{i}^{2}\right].\end{equation}
 The center-of-mass motion is simply that of a harmonically trapped
particle of mass $M=Nm$, with well-known wave functions and spectrum
$E_{cm}=(n_{cm}+3/2)\hbar\omega$, where $n_{cm}=0,1,2...$ is a non-negative
integer. In the presence of interactions, the relative Hamiltonian
should be solved in conjunction with the Bethe-Peierls boundary condition,
Eq. (\ref{BP}).

\subsection{Two fermions in a 3D harmonic trap}

Let us first briefly revisit the two-fermion problem in a harmonic
trap, where the relative Schrödinger equation becomes \begin{equation}
\left[-\frac{\hbar^{2}}{2\mu}{\bf \nabla}_{{\bf r}}^{2}+\frac{1}{2}\mu\omega^{2}r^{2}\right]\psi_{2b}^{rel}({\bf r})=E_{rel}\psi_{2b}^{rel}({\bf r}),\label{hamiRel2e}\end{equation}
 where two fermions with unlike spins do not stay at the same position
($r>0$). Here, we have re-defined ${\bf r}=\sqrt{2}{\bf r}_{2}$
and have used a reduced mass $\mu=m/2$. It is clear that only the
$l=0$ subspace of the relative wave function is affected by the $s$-wave
contact interaction. According to the Bethe-Peierls boundary condition,
as $r\rightarrow0$ the relative wave function should take the form,
$\psi_{2b}^{rel}(r)\rightarrow(1/r-1/a)$, or satisfy, $\partial\left(r\psi_{2b}^{rel}\right)/\partial r=-\left(r\psi_{2b}^{rel}\right)/a$.
The two-fermion problem in a harmonic trap was first solved by Busch
and coworkers \cite{busch}. Here, we present a simple physical interpretation
of the solution.

The key point is that, regardless of the boundary condition, there
are {\em two} types of general solutions of the relative Schrödinger
equation (\ref{hamiRel2e}) in the $l=0$ subspace, $\psi_{2b}^{rel}(r)\propto\exp(-r^{2}/2d^{2})f(r/d)$.
Here the function $f(x)$ can either be the first kind of Kummer confluent
hypergeometric function $_{1}F_{1}$ or the second kind of Kummer
confluent hypergeometric function $U$. We have taken $d=\sqrt{\hbar/\mu\omega}$
as the characteristic length scale of the trap. In the absence of
interactions, the first Kummer function gives rise to the standard
wave function of 3D harmonic oscillators. With interactions, however,
we have to choose the second Kummer function $U$, since it diverges
as $1/r$ at origin and thus satisfies the Bethe-Peierls boundary
condition. 

Therefore, the (un-normalized) relative wave function and relative
energy should be rewritten as, \begin{equation}
\psi_{2b}^{rel}(r;\nu)=\Gamma(-\nu)U(-\nu,\frac{3}{2},\frac{r^{2}}{d^{2}})\exp(-\frac{r^{2}}{2d^{2}}),\end{equation}
 and \begin{equation}
E_{rel}=(2\nu+\frac{3}{2})\hbar\omega,\label{spectrumRel2e}\end{equation}
 respectively. Here, $\Gamma$ is the Gamma function, the real number
$\nu$ plays the role of a quantum number and should be determined
by the boundary condition, $\lim_{r\rightarrow0}\partial\left(r\psi_{2b}^{rel}\right)/\partial r=-\left(r\psi_{2b}^{rel}\right)/a$.
By examining the short range behavior of the second Kummer function
$U(-\nu,3/2,x)$, this leads to the familiar equation for energy levels,
\begin{equation}
\frac{2\Gamma(-\nu)}{\Gamma(-\nu-1/2)}=\frac{d}{a}.\label{seRel2e}\end{equation}
 In Fig. 1, we give the resulting energy spectrum as a function of
the dimensionless interaction strength $d/a$.

The spectrum is easy to understand. At infinitely small scattering
length $a\rightarrow0^{-}$, $\nu(a=0^{-})=n_{rel}$ ($n_{rel}=0,1,2...$),
which recovers the spectrum in the non-interacting limit. With increasingly
attractive interactions, the energies decrease. In the unitarity (resonance)
limit where the scattering length diverges, $a\rightarrow\pm\infty$,
we find that $\nu(a=\pm\infty)=n_{rel}-1/2$. As the attraction increases
further, the scattering length becomes positive and decreases in magnitude.
We then observe two distinct types of behavior: the ground state is
a {\em molecule} of size $a$, whose energy diverges asymptotically
as $-\hbar^{2}/ma^{2}$ as $a\rightarrow0^{+}$, while the excited
states may be viewed as two {\em repulsively} interacting fermions
with the same scattering length $a$. Their energies decrease to the
non-interacting values as $a\rightarrow0^{+}$.

In this two-body picture, a universal repulsively interacting Fermi
gas with zero-range interaction potentials may be realized on the
positive scattering length side of a Feshbach resonance for an attractive
interaction potential, provided that all two fermions with unlike
spins occupy the exited states or the upper branch of the two-body
energy spectrum.

\begin{figure}
\begin{centering}
\includegraphics[clip,width=0.35\textwidth]{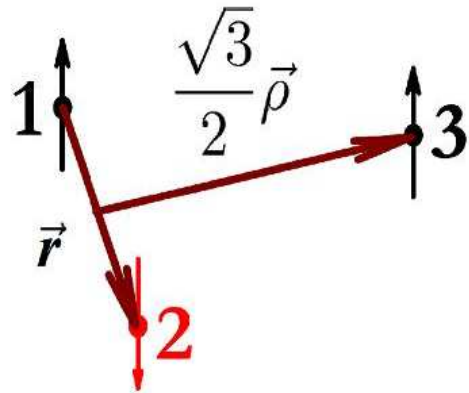} 
\par\end{centering}

\caption{(Color online) Configuration of three interacting fermions, two spin-up
and one spin-down.}

\label{fig2} 
\end{figure}

\subsection{Three fermions in a 3D harmonic trap}

Let us turn to the three fermion case by considering two spin-up fermions
and one spin-down fermion, i.e., the $\uparrow\downarrow\uparrow$
configuration shown in Fig. 2. The relative Hamiltonian can be written
as \cite{wernerpra,footnote}, \begin{equation}
{\cal H}_{rel}=\frac{\hbar^{2}}{2\mu}{\bf \left(\nabla_{r}^{2}+\nabla_{\rho}^{2}\right)}+\frac{1}{2}\mu\omega^{2}\left(r^{2}+\rho{\bf ^{2}}\right),\label{hamiRel3e}\end{equation}
 where we have redefined the Jacobi coordinates ${\bf r}=\sqrt{2}{\bf r_{2}}$
and ${\bf \rho}=\sqrt{2}{\bf r_{3}}$, which measure the distance
between the particle 1 and 2 (i.e., pair), and the distance from the
particle 3 to the center-of-mass of the pair, respectively.

\subsubsection{General exact solutions}

Inspired by the two-fermion solution, it is readily seen that the
relative wave function of the Hamiltonian (\ref{hamiRel3e}) may be
expanded into products of two Kummer confluent hypergeometric functions.
Intuitively, we may write down the following ansatz \cite{liu2009},
\begin{equation}
\psi_{3b}^{rel}\left({\bf r},{\bf \rho}\right)=\left(1-{\cal P}_{13}\right)\chi\left({\bf r},{\bf \rho}\right),\label{wfRel3e}\end{equation}
 where, \begin{equation}
\chi\left({\bf r},{\bf \rho}\right)=\sum\limits _{n}a_{n}\psi_{2b}^{rel}(r;\nu_{l,n})R_{nl}\left(\rho\right)Y_{l}^{m}\left(\hat{\rho}\right).\end{equation}
 The two-body relative wave function $\psi_{2b}^{rel}(r;\nu_{l,n})$
with energy $(2\nu_{l,n}+3/2)\hbar\omega$ describes the motion of
the paired particles 1 and 2, and the wave function $R_{nl}\left(\rho\right)Y_{l}^{m}\left(\hat{\rho}\right)$
with energy $(2n+l+3/2)\hbar\omega$ gives the motion of particle
3 relative to the pair. Here, $R_{nl}\left(\rho\right)$ is the standard
radial wave function of a 3D harmonic oscillator and $Y_{l}^{m}\left(\hat{\rho}\right)$
is the spherical harmonic. Owing to the rotational symmetry of the
relative Hamiltonian (\ref{hamiRel3e}), it is easy to see that the
relative angular momenta $l$ and $m$ are good quantum numbers. The
value of $\nu_{l,n}$ is uniquely determined from energy conservation,
\begin{equation}
E_{rel}=\left[(2\nu_{l,n}+3/2)+(2n+l+3/2)\right]\hbar\omega,\end{equation}
 for a given relative energy $E_{rel}$. It varies with the index
{}$n$ at a given angular momentum $l$. Finally, ${\cal P}_{13}$
is an exchange operator for particles 1 and 3, which ensures the correct
exchange symmetry of the relative wave function due to Fermi exclusion
principle, i.e., ${\cal P}_{13}\chi\left({\bf r},{\bf \rho}\right)=\chi\left({\bf r}/2+\sqrt{3}{\bf \rho}/2,{\bf \sqrt{3}r}/2-{\bf \rho}/2\right)$.
The relative energy $E_{rel}$ together with the expansion coefficient
$a_{n}$ should be determined by the Bethe-Peierls boundary condition,
i.e., $\lim_{r\rightarrow0}[\partial r\psi_{3b}^{rel}\left({\bf r},{\bf \rho}\right)]/\partial r=-[r\psi_{3b}^{rel}\left({\bf r},{\bf \rho}\right)]/a$.
We note that the second Bethe-Peierls boundary condition in case of
particle 2 approaching particle 3 is satisfied automatically due to
the exchange operator acting on the relative wave function.

By writing $\chi\left({\bf r},{\bf \rho}\right)=\phi(r,\rho)Y_{l}^{m}\left(\hat{\rho}\right)$,
the Bethe-Peierls boundary condition takes the form ($r\rightarrow0$),
\begin{equation}
-\frac{1}{a}\left[r\phi(r,\rho)\right]=\frac{\partial\left[r\phi(r,\rho)\right]}{\partial r}-\left(-1\right)^{l}\phi(\frac{\sqrt{3}\rho}{2},\frac{\rho}{2}).\label{BP3e}\end{equation}
 Using the asymptotic behavior of the second kind of Kummer function,
$\lim_{x\rightarrow0}\Gamma\left(-\nu_{l,n}\right)U(-\nu_{l,n},3/2,x^{2})=\sqrt{\pi}/x-2\sqrt{\pi}\Gamma\left(-\nu_{l,n}\right)/\Gamma\left(-\nu_{l,n}-1/2\right)$,
it is easy to show that in the limit of $r\rightarrow0$, \begin{equation}
-\frac{1}{a}\left[r\phi(r,\rho)\right]=-\frac{\sqrt{\pi}}{a}\sum\limits _{n}a_{n}R_{nl}\left(\rho\right),\label{BP3e1}\end{equation}
 and \begin{equation}
\frac{\partial\left[r\phi(r,\rho)\right]}{\partial r}=-\sqrt{\pi}\sum\limits _{n}a_{n}R_{nl}\left(\rho\right)\frac{2\Gamma\left(-\nu_{l,n}\right)}{\Gamma\left(-\nu_{l,n}-1/2\right)}.\label{BP3e2}\end{equation}

Thus, the Bethe-Peierls boundary condition becomes, \begin{equation}
\sum\limits _{n}a_{n}\left[B_{n}R_{nl}\left(\rho\right)-R_{nl}\left(\frac{\rho}{2}\right)\psi_{2b}^{rel}(\frac{\sqrt{3}\rho}{2};\nu_{l,n})\right]=0,\end{equation}
 where \begin{equation}
B_{n}=\left(-1\right)^{l}\sqrt{\pi}\left[\frac{d}{a}-\frac{2\Gamma\left(-\nu_{l,n}\right)}{\Gamma\left(-\nu_{l,n}-1/2\right)}\right].\end{equation}
 Projecting onto the orthogonal and complete set of basis functions
$R_{nl}\left(\rho\right)$, we find that a secular equation, \begin{equation}
\frac{2\Gamma(-\nu_{l,n})}{\Gamma(-\nu_{l,n}-1/2)}a_{n}+\frac{(-1)^{l}}{\sqrt{\pi}}\sum\limits _{n^{\prime}}C_{nn^{\prime}}a_{n^{\prime}}=\left(\frac{d}{a}\right)a_{n},\label{seRel3e}\end{equation}
 where we have defined the matrix coefficient, \begin{equation}
C_{nn^{\prime}}\equiv\int\limits _{0}^{\infty}\rho^{2}d\rho R_{nl}\left(\rho\right)R_{n^{\prime}l}\left(\frac{\rho}{2}\right)\psi_{2b}^{rel}(\frac{\sqrt{3}\rho}{2};\nu_{l,n^{\prime}}),\end{equation}
 which arises from the exchange effect due to the operator ${\cal P}_{13}$.
In the absence of $C_{nn^{\prime}}$, the above secular equation describes
a three-fermion problem of a pair and a single particle, {\em un-correlated}
to each other. It then simply reduces to Eq. (\ref{seRel2e}), as
expected.

The secular equation (\ref{seRel3e}) was first obtained by Kestner
and Duan by solving the three-particle scattering problem using Green
function \cite{kestner}. To solve it, for a given scattering length
we may try different values of relative energy $E_{rel}$, implicit
via $\nu_{l,n}$. However, it turns out to be more convenient to diagonalize
the matrix ${\bf A}=\{A_{nn^{\prime}}\}$ for a given relative energy,
where \begin{equation}
A_{nn^{\prime}}=\frac{2\Gamma(-\nu_{l,n})}{\Gamma(-\nu_{l,n}-1/2)}\delta_{nn^{\prime}}+\frac{(-1)^{l}}{\sqrt{\pi}}C_{nn^{\prime}}.\label{amat}\end{equation}
 The eigenvalues of the matrix ${\bf A}$ then gives all the possible
values of $d/a$ for a particular relative energy. We finally invert
$a(E_{rel})$ to obtain the relative energy as a function of the scattering
length. Numerically, we find that the matrix ${\bf A}$ is symmetric
and thus the standard diagonalization algorithm can be used. We outline
the details of the numerical calculation of Eq. (\ref{amat}) in the
Appendix A.

\subsubsection{Exact solutions in the unitarity limit}

In the unitarity limit with infinitely large scattering length, $a\rightarrow\infty$,
we may obtain more physical solutions using hyperspherical coordinates,
as shown by Werner and Castin \cite{wernerprl,footnote}. By defining
a hyperradius $R=\sqrt{(r^{2}+\rho^{2})/2}$ and hyperangles $\vec{\Omega}=(\alpha,\hat{r},\hat{\rho})$,
where $\alpha=\arctan(r/\rho)$ and $\hat{r}$ and $\hat{\rho}$ are
respectively the unit vector along ${\bf r}$ and ${\bf \rho}$, we
may write \cite{wernerprl,footnote}, \begin{equation}
\psi_{3b}^{rel}\left(R,\vec{\Omega}\right)=\frac{F\left(R\right)}{R}\left(1-{\cal P}_{13}\right)\frac{\varphi\left(\alpha\right)}{\sin\left(2\alpha\right)}Y_{l}^{m}\left(\hat{\rho}\right),\label{wfRel3eHyper}\end{equation}
 to decouple the motion in the hyperradius and hyperangles for given
relative angular momenta $l$ and $m$ . It leads to the following
decoupled Schrödinger equations \cite{footnote}, \begin{equation}
-F^{\prime\prime}-\frac{1}{R}F^{\prime}+\left(\frac{s_{l,n}^{2}}{R^{2}}+\omega^{2}R^{2}\right)F=2E_{rel}F,\label{hyperradiusEq}\end{equation}
 and \begin{equation}
-\varphi^{\prime\prime}\left(\alpha\right)+\frac{l\left(l+1\right)}{\cos^{2}\alpha}\varphi\left(\alpha\right)=s_{l,n}^{2}\varphi\left(\alpha\right),\label{hyperangleEq}\end{equation}
 where $s_{l,n}^{2}$ is the eigenvalue for the $n$-th wave function
of the hyperangle equation.

For three-fermions, $s_{l,n}^{2}$ is always positive. Therefore,
the hyperradius equation (\ref{hyperradiusEq}) can be interpreted
as a Schrödinger equation for a fictitious particle of mass unity
moving in two dimensions in an effective potential $(s_{l,n}^{2}/R^{2}+\omega^{2}R^{2})$
with a bounded wave function $F(R)$. The resulting spectrum is \cite{wernerprl,footnote}
\begin{equation}
E_{rel}=\left(2q+s_{l,n}+1\right)\hbar\omega,\label{energyRel3eHyper}\end{equation}
 where the good quantum number $q$ labels the number of nodes in
the hyperradius wave function.

The eigenvalue $s_{l,n}$ should be determined by the Bethe-Peierls
boundary condition, which in hyperspherical coordinates takes the
from \cite{wernerprl,footnote}, \begin{equation}
\varphi^{\prime}\left(0\right)-(-1)^{l}\frac{4}{\sqrt{3}}\varphi\left(\frac{\pi}{3}\right)=0.\label{BP3eHyper}\end{equation}
 In addition, we need to impose the boundary condition $\varphi\left(\pi/2\right)=0$,
since the relative wave function (\ref{wfRel3eHyper}) should not
be singular at $\alpha=\pi/2$. The general solution of the hyperangle
equation (\ref{hyperangleEq}) satisfying $\varphi\left(\pi/2\right)=0$
is given by, \begin{equation}
\varphi\propto x^{l+1}{}_{2}F_{1}\left(\frac{l+1-s_{l,n}}{2},\frac{l+1+s_{l,n}}{2},l+\frac{3}{2};x^{2}\right),\label{wfHyperangleEq}\end{equation}
 where $x=\cos(\alpha)$ and $_{2}F_{1}$ is the hypergeometric function.
In the absence of interactions, the Bethe-Peierls boundary condition
(\ref{BP3eHyper}) should be replaced by $\varphi\left(0\right)=0$,
since the relative wave function (\ref{wfRel3eHyper}) should not
be singular at $\alpha=0$ either. As $\varphi\left(0\right)=\Gamma(l+3/2)\Gamma(1/2)/[\Gamma((l+2+s_{l,n})/2)\Gamma((l+2-s_{l,n})/2)]$,
this boundary condition leads to $[l+2-s_{l,n}^{(1)}]/2=-n$, or $s_{l,n}^{(1)}=2n+l+2$,
where $n=0,1,2,...$ is a non-negative integer and we have used the
superscript {}``$1$'' to denote a non-interacting system. However,
a spurious solution occurs when $l=0$ and $n=0$, for which $s_{l,n}^{(1)}=2$,
$\varphi(\alpha)=\sin(2\alpha)/2$ and thus, the symmetry operator
$(1-{\cal P}_{13})$ gives a vanishing relative wave function in Eq.
(\ref{wfRel3eHyper}) that should be discarded \cite{footnote}. We
conclude that for three non-interacting fermions, \begin{equation}
s_{l,n}^{(1)}=\left\{ \begin{array}{ll}
2n+4, & l=0\\
2n+l+2, & l>0\end{array}\right..\label{non_interacting_sln}\end{equation}
 For three interacting fermions, we need to determine $s_{l,n}$ by
substituting the general solution (\ref{wfHyperangleEq}) into the
Bethe-Peierls boundary condition (\ref{BP3eHyper}). In the Appendix
B, we describe how to accurately calculate $s_{l,n}$. In the boundary
condition Eq. (\ref{BP3eHyper}), the leading effect of interactions
is carried by $\varphi^{\prime}\left(0\right)$ and therefore, $\varphi^{\prime}\left(0\right)=0$
determines the asymptotic values of $s_{l,n}$ at large momentum $l$
or $n$. This gives rise to $(l+1-\bar{s}_{l,n})/2=-n$, or, \begin{equation}
\bar{s}_{l,n}=\left\{ \begin{array}{ll}
2n+3, & l=0\\
2n+l+1, & l>0\end{array}\right.,\label{asymptotic_sln}\end{equation}
 where we have used a bar to indicate the asymptotic results. By comparing
Eqs. (\ref{non_interacting_sln}) and (\ref{asymptotic_sln}), asymptotically
the attractive interaction will reduce $s_{l,n}$ by a unity.

\begin{figure}
\begin{centering}
\includegraphics[clip,width=0.45\textwidth]{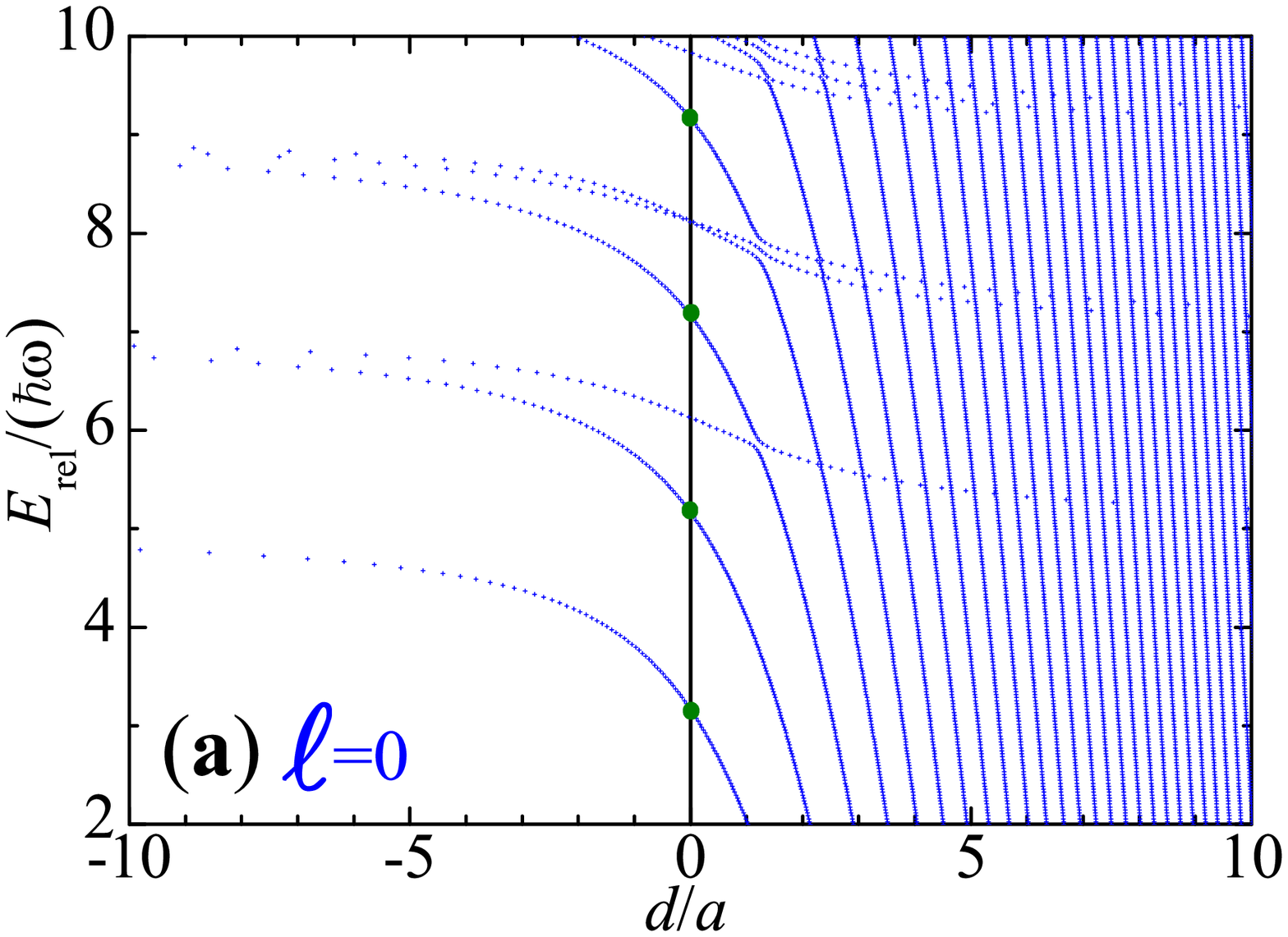} 
\par\end{centering}

\begin{centering}
\includegraphics[width=0.45\textwidth]{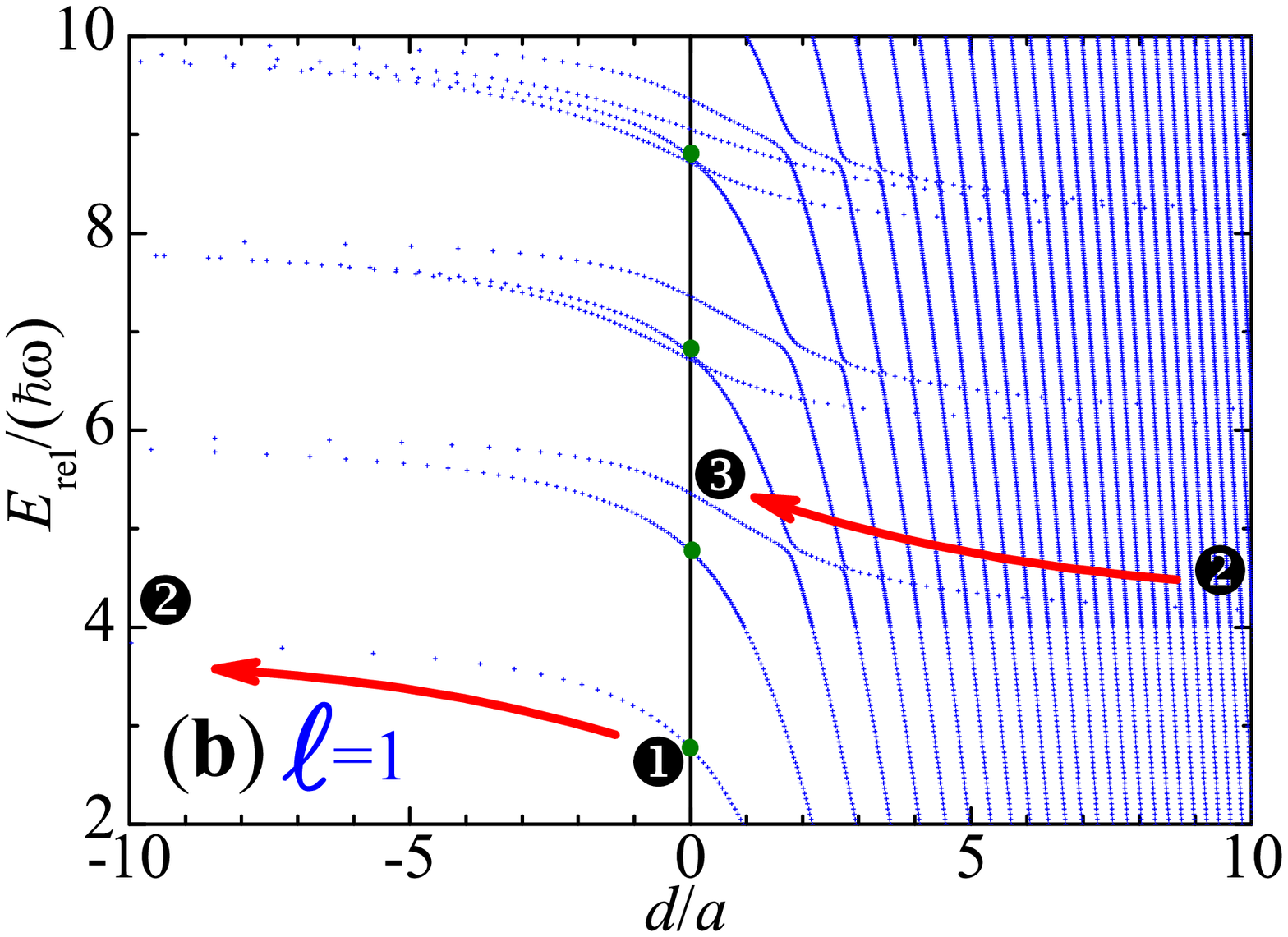} 
\par\end{centering}

\begin{centering}
\includegraphics[width=0.45\textwidth]{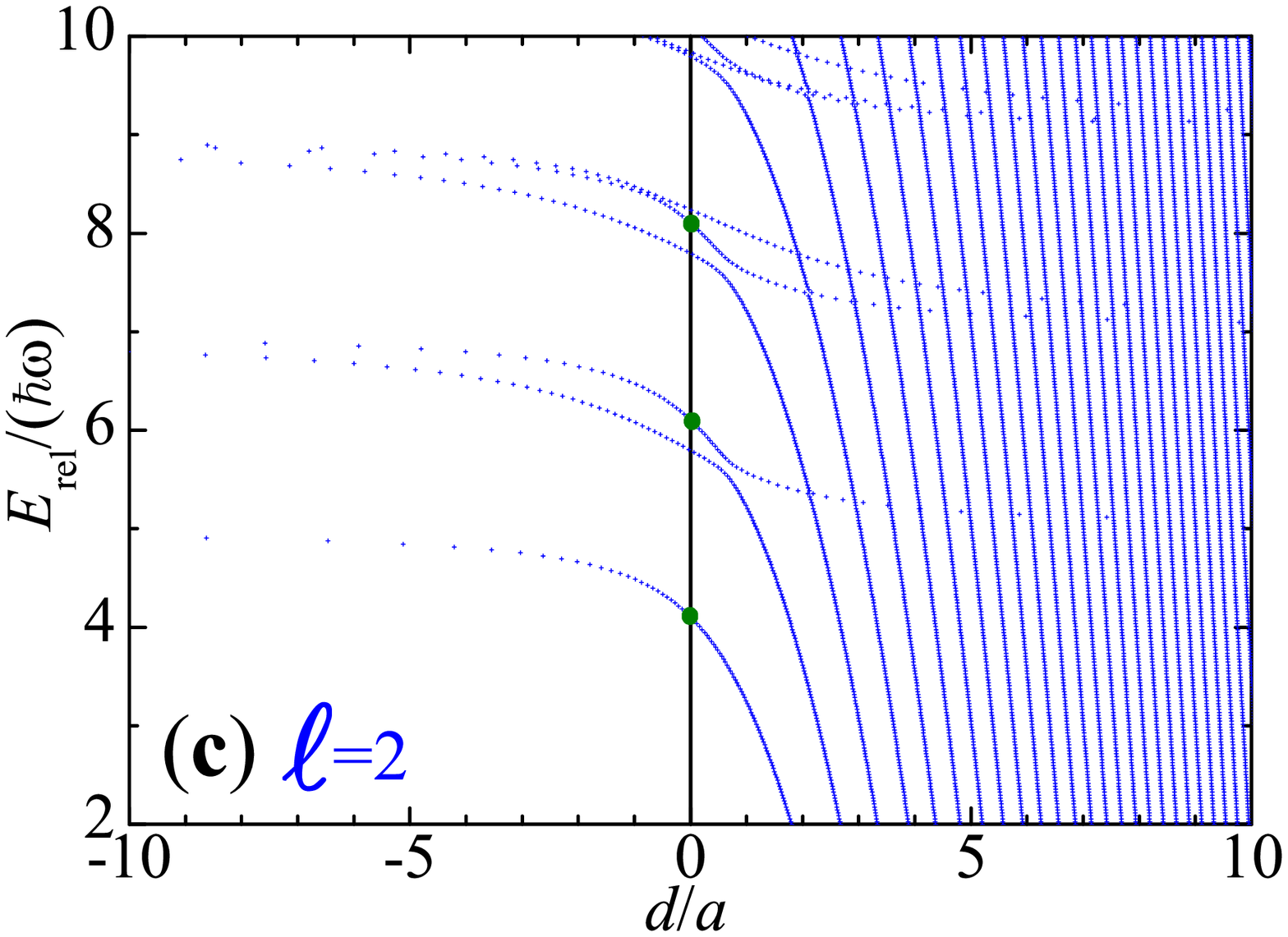} 
\par\end{centering}

\caption{(Color online) Relative energy spectrum of three interacting fermions
at different subspaces or relative angular momenta $l$. On the positive
scattering length (BEC) side of the resonance, there are two types
of energy levels: one (is vertical and) diverges with decreasing the
scattering length $a$ and the other (is horizontal) converges to
the non-interacting spectrum. The latter may be viewed as the energy
spectrum of three repulsively interacting fermions. In analogy with
the two-fermion case, we show in the ground state subspace ($l=1$)
the ground state energy level of the repulsive three-fermion system
(i.e, from point 2 to 3), as well as the ground state energy level
of the attractive three-fermion system for $a<0$ (i.e., from the
point 1 to 2). In the unitarity limit, we show by the circles the
energy levels that should be excluded when we identify the energy
spectrum for infinitely large repulsive interactions.}

\label{fig3} 
\end{figure}

\subsubsection{Energy spectrum of three interacting fermions}

We have numerically solved both the general exact solution (\ref{wfRel3e})
along the BEC-BCS crossover and the exact solution (\ref{wfRel3eHyper})
in the unitarity limit. In the latter unitary case, the accuracy of
results can be improved to arbitrary precision by using suitable mathematical
software, described in Appendix B. Fig. 3 reports the energy spectrum
of three interacting fermions with increasingly attractive interaction
strength at three total relative angular momenta, $l=0$, $1$, and
$2$. For a given scattering length, we typically calculate several
ten thousand energy levels (i.e., $E_{rel}<(l+256)\hbar\omega$) in
each different subspace. To construct the matrix ${\bf A}$, Eq. (\ref{amat}),
we have kept a maximum value of $n_{\max}=128$ in the functions $R_{nl}\left(\rho\right)$.
Using the accurate spectrum in the unitarity limit as a benchmark,
we estimate that the typical relative numerical error of energy levels
is less than $10^{-6}$. We have found a number of nontrivial features
in the energy spectrum.

The spectrum on the BCS side is relatively simple. It can be understood
as a non-interacting spectrum at $d/a\rightarrow-\infty$, in which
$E_{rel}=(2Q+3)\hbar\omega$ at $l=0$ and $E_{rel}=(2Q+l+1)\hbar\omega$
at $l\geq1$, with a positive integer $Q=1,2,3,...$ that denotes
also the degeneracy of the energy levels. The attractive interactions
reduce the energies and at the same time lift the degeneracy. Above
the resonance or unitary point of $d/a=0$, however, the spectrum
becomes much more complicated.

There are a group of nearly {\em vertical} energy levels that diverge
towards the BEC limit of $d/a\rightarrow+\infty$. From the two-body
relative energy spectrum in Fig. 1, we may identify these as energy
states containing a molecule of size $a$ and a fermion. For a given
scattering length, these nearly vertical energy level differ by about
$2\hbar\omega$, resulting from the motion of the fermion relative
to the molecule. In addition to the nearly vertical energy levels,
most interestingly, we observe also some nearly {\em horizontal}
energy levels, which converge to the non-interacting spectrum in the
BEC limit. In analogy with the two-body case, we may identify these
horizontal levels as the energy spectrum of three {\em repulsively}
interacting fermions. We show explicitly in Fig. 3b the ground state
level of three repulsively interacting fermions, which increases in
energy from the point 2 to 3 with increasing scattering length from
$a=0^{+}$ to $a=+\infty$. For comparison, we also show the ground
state level of three attractively interacting fermions at a negative
scattering length, which decreases in energy from the point 2 to 1
with increasing absolute value of $a$.

This identification of energy spectrum for repulsive interactions,
however, is not as rigorous as in the two-body case. There are many
apparent avoided crossings between the vertical and horizontal energy
levels. Therefore, by changing a positive scattering length from the
BEC limit to the unitarity limit, three fermions initially at the
horizontal level may finally transition into a vertical level, provided
that the sweep of scattering length is sufficiently slow and adiabatic.
This leads to the conversion of fermionic atoms to bosonic molecules.
A detailed analysis of the loss rate of fermionic atoms as a function
of sweep rate may be straightforward obtained by applying the Landau-Zener
tunnelling model.

Let us now focus on the resonance case of most significant interest.
In Fig. 3, we show explicitly by green dots the vertical energy levels
in the unitarity limit. These levels should be excluded if we are
interested in the spectrum of repulsively interacting fermions. Amazingly,
for each given angular momentum, these energy levels form a regular
ladder with an exact energy spacing $2\hbar\omega$ \cite{wernerpra}.
Using the exact solution in the unitarity limit, Eq. (\ref{energyRel3eHyper}),
we may identify unambiguously that the energy ladder is given by,
\begin{equation}
E_{rel}=\left(2q+s_{l,0}+1\right)\hbar\omega.\label{energyRel3eLowest}\end{equation}
 Therefore, in the unitarity limit the lowest-order solution of the
hyperangle equation gives rise to the relative wave function of a
molecule and a fermion. Thus, it should be discarded when considering
three resonantly interacting fermions with an effective repulsive
interaction.

This observation immediately leads to the ground state energy of three
repulsively interacting fermions, \begin{equation}
E_{gs}^{\uparrow\downarrow\uparrow}=\left(s_{1,1}+1\right)\hbar\omega+1.5\hbar\omega\simeq6.858249309\hbar\omega,\end{equation}
 including the zero-point energy of the center-of-mass motion, $1.5\hbar\omega$.
This ground state energy is higher than that of three fully polarized
fermions, which is, \begin{equation}
E_{gs}^{\uparrow\uparrow\uparrow}=1.5\hbar\omega+2.5\hbar\omega+2.5\hbar\omega=6.5\hbar\omega.\end{equation}
 Thus, in the presence of repulsive interactions, the ground state
of three fully polarized fermions is unstable with respect to a single
spin-flip, suggestive of an itinerant ferromagnetic transition at
a certain scattering length for three fermions.

\section{Quantum virial expansion for thermodynamics}

The few-particle solutions presented above can provide information
about the high temperature thermodynamics of many-body systems, through
a quantum virial expansion of the grand thermodynamic potential \cite{liu2009,ho2004b}.
In the grand canonical ensemble, the thermodynamic potential is given
by, \begin{equation}
\Omega=-k_{B}T\ln{\cal Z},\end{equation}
 where $k_{B}$ is the Boltzmann constant and \begin{equation}
{\cal Z}=\text{Tr}\exp\left[-\left({\cal H}-\mu{\cal N}\right)/k_{B}T\right]\end{equation}
 is the grand partition function. We may rewrite this in terms of
the partition function of clusters, \begin{equation}
Q_{n}=\text{Tr}_{n}\left[\exp\left(-{\cal H}/k_{B}T\right)\right],\end{equation}
 where the integer $n$ denotes the number of particles in the cluster
and the trace Tr$_{n}$ is taken over $n$-particle states with a
proper symmetry. The partition function of clusters $Q_{n}$ can be
calculated using the complete solutions of a $n$-particles system.
The grand partition function is then written as \begin{equation}
{\cal Z}=1+zQ_{1}+z^{2}Q_{2}+\cdots,\end{equation}
 where $z=\exp\left(\mu/k_{B}T\right)$ is the fugacity. At high temperatures,
it is well-known that the chemical potential $\mu$ diverges to $-\infty$,
so the fugacity would be very small, $z\ll1$. We can then expand
the high-temperature thermodynamic potential in powers of the small
parameter $z$, \begin{equation}
\Omega=-k_{B}TQ_{1}\left[z+b_{2}z^{2}+\cdots+b_{n}z^{n}+\cdots\right],\end{equation}
 where $b_{n}$ may be referred to as the $n$-th (virial) expansion
coefficient. It is readily seen that, \begin{eqnarray}
b_{2} & = & \left(Q_{2}-Q_{1}^{2}/2\right)/Q_{1},\\
b_{3} & = & \left(Q_{3}-Q_{1}Q_{2}+Q_{1}^{3}/3\right)/Q_{1},\ etc.\end{eqnarray}
 These equations present a general definition of the quantum virial
expansion and are applicable to both homogeneous and trapped systems.
The determination of the $n$-th virial coefficient requires knowledge
of up to the $n$-body problem.

In practice, it is convenient to concentrate on the interaction effects
only. We thus consider the difference $\Delta b_{n}\equiv b_{n}-b_{n}^{(1)}$
and $\Delta Q_{n}\equiv Q_{n}-Q_{n}^{(1)}$, where the superscript
{}``$1$'' denotes the non-interacting systems. For the second and
third virial coefficient, we shall calculate \begin{equation}
\Delta b_{2}=\Delta Q_{2}/Q_{1}\end{equation}
 and \begin{equation}
\Delta b_{3}=\Delta Q_{3}/Q_{1}-\Delta Q_{2}.\end{equation}

\subsection{Non-interacting virial coefficients}

The background non-interacting virial coefficients can be conveniently
determined by the non-interacting thermodynamic potential. For a \emph{homogeneous}
two-component Fermi gas, this takes the form, \begin{equation}
\Omega_{\hom}^{(1)}=-V\frac{2k_{B}T}{\lambda^{3}}\frac{2}{\sqrt{\pi}}\int\limits _{0}^{\infty}t^{1/2}\ln\left(1+ze^{-t}\right)dt,\label{non_interacting_Omega_Homo}\end{equation}
 where $\lambda\equiv[2\pi\hbar^{2}/(mk_{B}T)]^{1/2}$ is the thermal
wavelength and $Q_{1,\hom}=2V/\lambda^{3}$. This leads to \begin{equation}
b_{n,\hom}^{(1)}=\frac{\left(-1\right)^{n+1}}{n^{5/2}}.\end{equation}
 Hereafter, we use the subscript {}``hom'' to denote the quantity
in the homogeneous case, otherwise, by default we refer to a trapped
system. 

For a harmonically trapped Fermi gas, the non-interacting thermodynamic
potential in the semiclassical limit (neglecting the discreteness
of the energy spectrum) is, \begin{equation}
\Omega^{(1)}=-\frac{2\left(k_{B}T\right)^{4}}{\left(\hbar\omega\right)^{3}}\frac{1}{2}\int\limits _{0}^{\infty}t^{2}\ln\left(1+ze^{-t}\right)dt,\label{non_interacting_Omega_Trap}\end{equation}
 where $Q_{1}=2\left(k_{B}T\right)^{3}/\left(\hbar\omega\right)^{3}$.
Taylor-expanding in powers of $z$ gives rise to \begin{equation}
b_{n}^{(1)}=\frac{\left(-1\right)^{n+1}}{n^{4}}.\end{equation}
 We note that the non-interacting virial coefficients in the homogeneous
case and trapped case are related by, \begin{equation}
b_{n}^{(1)}=\frac{b_{n,\hom}^{(1)}}{n^{3/2}}.\label{non_bn_Trap_vs_Homo}\end{equation}

\subsection{Second virial coefficient in a harmonic trap}

We now calculate the second virial coefficient of a trapped interacting
Fermi gas. In a harmonic trap, the oscillator length $d$ provides
a large length scale, compared to the thermal wavelength $\lambda$.
Alternatively, we may use $\tilde{\omega}=\hbar\omega/k_{B}T\ll1$
to characterize the intrinsic length scale relative to the trap. All
the virial coefficients and cluster partition functions in harmonic
traps therefore depend on the small parameter $\tilde{\omega}$. We
shall be interested in a universal regime with vanishing $\tilde{\omega}$,
in accord with the large number of atoms in a real experiment.

To obtain $\Delta b_{2}$, we consider separately $\Delta Q_{2}$
and $Q_{1}$. The single-particle partition function $Q_{1}$ is determined
by the single-particle spectrum of a 3D harmonic oscillator, $E_{nl}=(2n+l+3/2)\hbar\omega$.
We find that $Q_{1}=2/[\exp(+\tilde{\omega}/2)-\exp(-\tilde{\omega}/2)]^{3}\simeq2\left(k_{B}T\right)^{3}/\left(\hbar\omega\right)^{3}$.
The prefactor of two accounts for the two possible spin states of
a single fermion. In the calculation of $\Delta Q_{2}$, it is easy
to see that the summation over the center-of-mass energy gives exactly
$Q_{1}/2$. Using Eq. (\ref{spectrumRel2e}), we find that, \begin{equation}
\Delta b_{2}^{att}=\frac{1}{2}\sum_{\nu_{n}}\left[e^{-\left(2\nu_{n}+3/2\right)\tilde{\omega}}-e^{-\left(2\nu_{n}^{\left(1\right)}+3/2\right)\tilde{\omega}}\right],\label{db2}\end{equation}
 where the non-interacting $\nu_{n}^{\left(1\right)}=n$ ($n=0,1,2,...$)
and the superscript {}``att'' (or {}``rep'') means the coefficient
of an attractively (or repulsively) interacting Fermi gas. The second
virial coefficient of a trapped attractive Fermi gas in the BEC-BCS
crossover was given in Fig. 3a of Ref. \cite{liu2009}.

To consider the second virial coefficient of a repulsively interacting
Fermi gas, we shall restrict ourselves to a positive scattering length
and exclude the lowest ground state energy level in the summation
of the first term in Eq. (\ref{db2}), which corresponds to a bound
molecule.

\subsubsection{Unitarity limit}

At resonance with an infinitely large scattering length, the spectrum
is known exactly: $\nu_{n,\infty}=n-1/2$, giving rise to, \begin{equation}
\Delta b_{2,\infty}^{att}=\frac{1}{2}\frac{\exp\left(-\tilde{\omega}/2\right)}{\left[1+\exp\left(-\tilde{\omega}\right)\right]}=+\frac{1}{4}-\frac{1}{32}\tilde{\omega}^{2}+\cdots.\label{db2attTrap}\end{equation}
 For a repulsive Fermi gas in the unitarity limit, we shall discard
the lowest `molecular' state with $\nu_{0,\infty}=-1/2$ and therefore,
\begin{equation}
\Delta b_{2,\infty}^{rep}=\frac{1}{2}\frac{\exp\left(-\tilde{\omega}/2\right)}{\left[1+\exp\left(+\tilde{\omega}\right)\right]}=-\frac{1}{4}+\frac{\tilde{\omega}}{4}+\cdots.\label{db2repTrap}\end{equation}
 The term $\tilde{\omega}^{2}$ or $\tilde{\omega}$ in Eqs. (\ref{db2attTrap})
and (\ref{db2repTrap}) is {\em nonuniversal} and is negligibly
small for a cloud with a large number of atoms. We therefore obtain
the universal second virial coefficients: $\Delta b_{2,\infty}^{att}=+1/4$
and $\Delta b_{2,\infty}^{rep}=-1/4$, which are temperature independent.

\subsection{Third virial coefficient in a harmonic trap}

The calculation of the third virial coefficient, which is given by
$\Delta b_{3}=\Delta Q_{3}/Q_{1}-\Delta Q_{2}$, is more complicated.
Either the term $\Delta Q_{3}/Q_{1}$ or $\Delta Q_{2}$ diverges
as $\tilde{\omega}\rightarrow0$, but the leading divergences cancel
with each other. In the numerical calculation, we have to carefully
separate the leading divergent term and calculate them analytically.
It is readily seen that the spin states of $\uparrow\downarrow\uparrow$
and $\downarrow\uparrow\downarrow$ configurations contribute equally
to $Q_{3}$. The term $Q_{1}$ in the denominators is canceled exactly
by the summation over the center-of-mass energy. We thus have \begin{equation}
\Delta Q_{3}/Q_{1}=[\sum\exp(-E_{rel}/k_{B}T)-\sum\exp(-E_{rel}^{(1)}/k_{B}T)]\,.\end{equation}

To proceed, it is important to analyze analytically the behavior of
$E_{rel}$ at high energies. For this purpose, we introduce a relative
energy $\bar{E}_{rel}$, which is the solution of Eq. (\ref{amat})
in the absence of the exchange term $C_{nm}$, and can be constructed
directly from the two-body relative energy. In the subspace with a
total relative momentum $l$, it takes the form, \begin{equation}
\bar{E}_{rel}=\left(2n+l+3/2\right)\hbar\omega+(2\nu+3/2)\hbar\omega,\label{ebar3e}\end{equation}
 where $\nu$ is the solution of the two-body spectrum of Eq. (\ref{seRel2e}).
At high energies the full spectrum $E_{rel}$ approaches asymptotically
to $\bar{E}_{rel}$ as the exchange effect becomes increasingly insignificant.
There is an important exception, however, occurring at zero total
relative momentum $l=0$. As mentioned earlier, the solution of $\bar{E}_{rel}$
at $n=0$ and $l=0$ is spurious and does not match any solution of
$E_{rel}$. Therefore, for the $l=0$ subspace, we require $n\geq1$
in Eq. (\ref{ebar3e}).

It is easy to see that if we keep the spurious solution in the $l=0$
subspace, the difference $[\sum\exp(-\bar{E}_{rel}/k_{B}T)-\sum\exp(-E_{rel}^{(1)}/k_{B}T)]$
is exactly equal to $\Delta Q_{2}$, since in Eq. (\ref{ebar3e})
the first part of spectrum is exactly identical to the spectrum of
center-of-mass motion. The spurious solution gives a contribution,
\begin{equation}
\sum_{\nu_{n}}\left[e^{-\left(2\nu_{n}+3\right)\tilde{\omega}}-e^{-\left(2\nu_{n}^{\left(1\right)}+3\right)\tilde{\omega}}\right]\equiv2e^{-3\tilde{\omega}/2}\Delta b_{2}^{att},\end{equation}
 which should be subtracted. Keeping this in mind, we finally arrive
at the following expression for the third virial coefficient of a
trapped Fermi gas with repulsive interactions: \begin{equation}
\Delta b_{3}^{att}=\sum\left[e^{-\frac{E_{rel}}{k_{B}T}}-e^{-\frac{\bar{E}_{rel}}{k_{B}T}}\right]-2e^{-3\tilde{\omega}/2}\Delta b_{2}^{att}.\end{equation}
 The summation is over all the possible relative energy levels $E_{rel}$
and their asymptotic values $\bar{E}_{rel}$. It is well-behaved and
converges at any scattering length. The third virial coefficient of
a trapped attractive Fermi gas in the BEC-BCS crossover was given
in Fig. 3b of Ref. \cite{liu2009}.

\subsubsection{Unitarity limit}

In the unitarity limit, it is more convenient to use the exact spectrum
given by Eq. (\ref{energyRel3eHyper}), where $s_{l,n}$ can be obtained
numerically to arbitrary accuracy and the non-interacting $s_{l,n}^{\left(1\right)}$
is given by Eq. (\ref{non_interacting_sln}). To control the divergence
problem, we shall use the same strategy as before and to approach
$s_{l,n}$ by using its asymptotic value $\bar{s}_{l,n}$ given in
Eq. (\ref{asymptotic_sln}).

Integrating out the $q$ degree of freedom and using Eq. (\ref{db2attTrap})
to calculate $\Delta Q_{2}$, we find that, \begin{equation}
\Delta b_{3,\infty}^{att}=\frac{e^{-\tilde{\omega}}}{1-e^{-2\tilde{\omega}}}\left[\sum_{l,n}\left(e^{-\tilde{\omega}s_{l,n}}-e^{-\tilde{\omega}\bar{s}_{l,n}}\right)+A\right],\end{equation}
 where $A$ is given by \begin{equation}
A=\sum_{l,n}\left(e^{-\tilde{\omega}\bar{s}_{l,n}}-e^{-\tilde{\omega}s_{l,n}^{\left(1\right)}}\right)-\frac{e^{-\tilde{\omega}}}{\left(1-e^{-\tilde{\omega}}\right)^{2}}.\end{equation}
 We note that for the summation, implicitly there is a prefactor $\left(2l+1\right)$,
accounting for the degeneracy of each subspace. The value of $A$
can then be calculated analytically, leading to, \begin{equation}
A=-e^{-\tilde{\omega}}\left(1-e^{-\tilde{\omega}}\right).\end{equation}
 We have calculated numerically $\sum_{l,n}(e^{-\tilde{\omega}s_{l,n}}-e^{-\tilde{\omega}\bar{s}_{l,n}})$
by imposing the cut-offs of $n<n_{\max}=512$ and $l<l_{\max}=512$.
We find that, \begin{equation}
\Delta b_{3,\infty}^{att}\simeq-0.06833960+0.038867\tilde{\omega}^{2}+\cdots.\end{equation}
 The numerical accuracy can be further improved by suitably enlarging
$n_{\max}$ and $l_{\max}$. For a Fermi gas with infinitely large
repulsions, we need to exclude the states involving a molecule. Thus,
in the calculation of $\Delta Q_{3}/Q_{1}$, we exclude the energy
levels associated with $s_{l,n=0}$, as given by Eq. (\ref{energyRel3eLowest}).
In the calculation of $\Delta Q_{2}$, we shall remove the lowest
two-body state with $\nu_{0,\infty}=-1/2$. In the end, we find that,
\begin{equation}
\Delta b_{3,\infty}^{rep}\simeq0.34976-0.77607\tilde{\omega}+\cdots.\end{equation}
 By neglecting the dependence on $\tilde{\omega}$ in the thermodynamic
limit, we obtain the universal third virial coefficients: \begin{eqnarray}
\Delta b_{3,\infty}^{att} & \simeq & -0.06833960,\\
\Delta b_{3,\infty}^{rep} & \simeq & 0.34976.\end{eqnarray}

\subsection{Unitary virial coefficients in homogeneous space}

We have so far studied the virial coefficients in a harmonic trap.
In the unitarity limit, there is a simple relation between the trapped
and homogeneous virial coefficient, as inspired by relation (\ref{non_bn_Trap_vs_Homo}).
This stems from the universal temperature independence of all virial
coefficients in the unitarity limit. In the thermodynamic limit, let
us consider the thermodynamic potential of a harmonic trapped Fermi
gas in the local density approximation $\Omega=\int d{\bf r}\Omega({\bf r})$,
where $\Omega({\bf r})$ is the local thermodynamic potential \begin{equation}
\Omega({\bf r})\propto z\left({\bf r}\right)+b_{2,\infty,\hom}z^{2}\left({\bf r}\right)+\cdots.\end{equation}
 Here, the local fugacity $z\left({\bf r}\right)=z\exp[-V({\bf r})/k_{B}T]$
is determined by the local chemical potential $\mu({\bf r})=\mu-V({\bf r})$.
On spatial integration, it is readily seen that the universal (temperature
independent) part of the trapped virial coefficient is, \begin{equation}
b_{n,\infty}=\frac{b_{n,\infty,\hom}}{n^{3/2}}.\end{equation}
 We therefore immediately obtain that the homogeneous second virial
coefficients in the unitarity limit are: \begin{eqnarray}
\Delta b_{2,\infty,\hom}^{att} & = & +\frac{1}{\sqrt{2}},\\
\triangle b_{2,\infty,\hom}^{rep} & = & -\frac{1}{\sqrt{2}},\end{eqnarray}
 and the homogeneous third virial coefficients are: \begin{eqnarray}
\Delta b_{3,\infty,\hom}^{att} & \simeq & -0.35501298,\\
\Delta b_{3,\infty,\hom}^{rep} & \simeq & +1.8174.\end{eqnarray}
 The homogeneous virial coefficients are therefore significantly larger
than their trapped counterparts. The factor of $n^{3/2}$ is clearly
due to the higher density of states in a harmonically trapped geometry.

\section{High-$T$ equation of state of a strongly interacting Fermi gas}

We are now ready to calculate the equation of states in the high temperature
regime, by using the thermodynamic potential \begin{equation}
\Omega_{\hom}=\Omega_{\hom}^{(1)}-V\frac{2k_{B}T}{\lambda^{3}}\left(\Delta b_{2,\hom}z^{2}+\cdots\right)\end{equation}
 and \begin{equation}
\Omega=\Omega^{(1)}-\frac{2\left(k_{B}T\right)^{4}}{\left(\hbar\omega\right)^{3}}\left(\Delta b_{2}z^{2}+\Delta b_{3}z^{3}+\cdots\right),\end{equation}
 respectively, for a homogeneous or a harmonically trapped Fermi gas.
Here, the non-interacting thermodynamic potentials are given by Eqs.
(\ref{non_interacting_Omega_Homo}) and (\ref{non_interacting_Omega_Trap}).
All the other thermodynamic quantities can be derived from the thermodynamic
potential by the standard thermodynamic relations, for example, $N=-\partial\Omega/\partial\mu$,
$S=-\partial\Omega/\partial T$, and then $E=\Omega+TS+\mu N$.

As an concrete example, let us focus on the unitarity limit in the
thermodynamic limit, which is of the greatest interest. The equations
of state are easy to calculate because of the temperature independence
of virial coefficients. It is also easy to check the well-known scaling
relation in the unitarity limit: $E=-3\Omega/2$ for a homogeneous
Fermi gas \cite{ho2004b} and $E=-3\Omega$ for a harmonically trapped
Fermi gas \cite{unitaritycmp}. The difference of the factor of two
arises from the fact that according to the virial theorem, in harmonic
traps the internal energy is exactly equal to the trapping potential
energy.

To be dimensionless, we take the Fermi temperature $T_{F}$ or Fermi
energy ($E_{F}=k_{B}T_{F}$) as the units for temperature and energy.
For a homogeneous or a harmonically trapped Fermi gas, the Fermi energy
is given by $E_{F}=\hbar^{2}(3\pi^{2}N/V)^{2/3}/2m$ and $E_{F}=(3N)^{1/3}\hbar\omega$,
respectively. In the actual calculations, we determine the number
of atoms $N$, the total entropy $S$, and the total energy $E$ at
given fugacity and a fixed temperature (i.e., $T=1$), and consequently
obtain the Fermi temperature $T_{F}$ and Fermi energy $E_{F}$. We
then plot the energy or energy per particle, $E/(NE_{F})$ and $S/(Nk_{B})$,
as a function of the reduced temperature $T/T_{F}$.

\begin{figure}[htp]

\begin{centering}
\includegraphics[clip,width=0.45\textwidth]{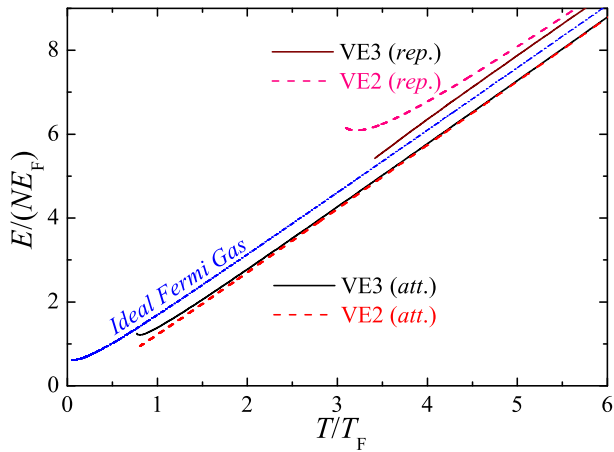} 
\par\end{centering}

\caption{(Color online) Energy per particle $E/(NE_{F})$ as a function of
reduce temperature $T/T_{F}$ for a homogeneous Fermi gas with infinitely
attractive and repulsive interactions. The predictions of quantum
virial expansion up to the second- and third-order are shown by solid
line and dashed line, respectively. For comparison, we plot the ideal
gas result by the dot-dashed line.}

\label{fig4} 
\end{figure}

\begin{figure}
\begin{centering}
\includegraphics[clip,width=0.45\textwidth]{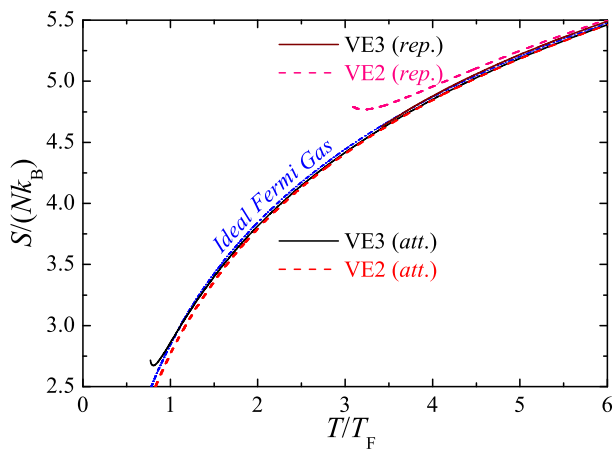} 
\par\end{centering}

\caption{(Color online) Entropy per particle $S/(NE_{F})$ as a function of
reduce temperature $T/T_{F}$ for a homogeneous Fermi gas with infinitely
attractive and repulsive interactions. The others are the same as
in Fig. 4.}

\label{fig5} 
\end{figure}

\subsection{Homogeneous equation of state}

We report in Figs. 4 and 5 the temperature dependence of energy and
entropy of a strongly attractively or repulsively interacting homogeneous
Fermi gas. The solid line and dashed line are the predictions of the
quantum virial expansion up to the third order (VE3) and second order
(VE2), respectively. For comparison, we also show the ideal gas result
by the thin dot-dashed line.

For a strongly attractively interacting Fermi gas, we observe that
the quantum virial expansion is valid down to the degeneracy temperature
$T_{F}$, where the predictions using the second-order or third-order
expansion do not greatly differ. We note that our prediction of the
third virial coefficient of a unitarity Fermi gas, $\Delta b_{3,\infty,\hom}^{att}\simeq-0.35501298$,
was experimentally confirmed to within $5\%$ relative accuracy in
the most recent thermodynamic measurement at ENS by Nascimbène and
co-workers \cite{ensexpt}.

However, for a strongly repulsively interacting Fermi gas, the applicability
of the quantum virial expansion is severely reduced: it seems to be
applicable only for $T>5T_{F}$. Below this characteristic temperature,
the difference between the second-order and third-order prediction
becomes very significant. This is partly due to the large absolute
value of the third virial coefficient, suggesting that in this case
the virial expansion converges very slowly.

\begin{figure}
\begin{centering}
\includegraphics[clip,width=0.45\textwidth]{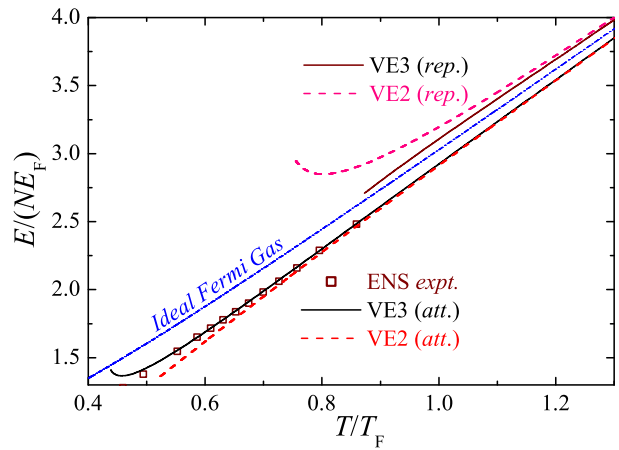} 
\par\end{centering}

\caption{(Color online) Energy per particle $E/(NE_{F})$ as a function of
reduced temperature $T/T_{F}$ for a trapped Fermi gas with infinitely
attractive and repulsive interactions. The predictions of quantum
virial expansion up to the second- and third-order are shown by solid
line and dashed line, respectively. For comparison, we plot the ideal
gas result by the dot-dashed line. We show also the experimental data
measured at ENS by empty squares for an attractive Fermi gas at unitarity
\cite{unitaritycmp,ensexpt}, which agree extremely well with the
prediction from quantum virial expansion.}

\label{fig6} 
\end{figure}

\begin{figure}
\begin{centering}
\includegraphics[clip,width=0.45\textwidth]{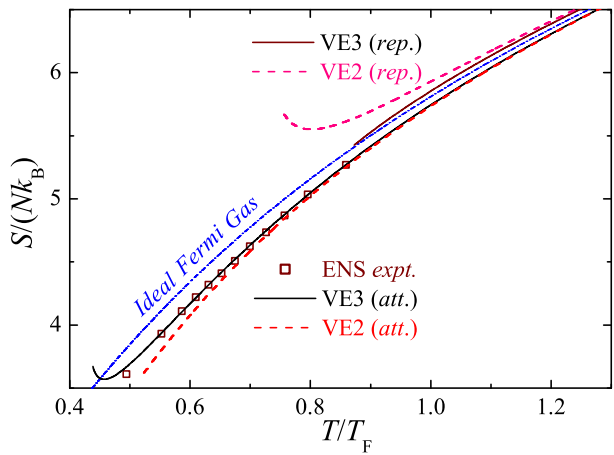} 
\par\end{centering}

\caption{(Color online) Entropy per particle $S/(NE_{F})$ as a function of
reduce temperature $T/T_{F}$ for a trapped Fermi gas with infinitely
attractive and repulsive interactions. The others are the same as
in Fig. 6.}

\label{fig7} 
\end{figure}

\subsection{Harmonically trapped equation of states}

We finally present in Figs. 6 and 7 the high-temperature expansion
prediction for the equation of state of a harmonically trapped Fermi
gas in the strongly interacting regime. Due to the significantly reduced
virial coefficients, the virial expansion in a trap has much broader
applicability. For a strongly attractively interacting Fermi gas,
it is now quantitatively applicable down to $0.5T_{F}$, as confirmed
by the precise experimental measurement at ENS (empty squares) \cite{unitaritycmp,ensexpt}.
At the same time, the virial expansion for a strongly repulsively
interacting gas seems to be qualitatively valid at $T>T_{F}$. At
this temperature, the energy of the repulsively interacting Fermi
gas is only marginally higher than the ideal, non-interacting energy.
Considering the large energy difference between a fully polarized
Fermi gas and a non-polarized Fermi gas (i.e., at the order of $NE_{F}$),
we conjecture that a strongly repulsively interacting Fermi gas does
not have itinerant ferromagnetism in the temperature regime where
the quantum virial expansion theory is applicable.

\section{Conclusions and outlook}

In conclusion, we have presented a complete set of exact solutions
for three attractively interacting fermions in a harmonic trap, with
either positive or negative scattering lengths. Firstly, we have outlined
the details of our previous studies on the quantum virial expansion
\cite{liu2009}, in particular the method for calculating the third
virial coefficient which was recently confirmed experimentally. In
addition, we have opened up the previously unexplored repulsively
interacting regime, and have presented a few-body perspective of itinerant
ferromagnetism. We have also studied the high-temperature thermodynamics
of a strongly repulsively interacting Fermi gas, by calculating its
second and third virial coefficients in the unitarity limit.

On the positive scattering length side of a Feshbach resonance, a
repulsively interacting Fermi gas is thought to occur by excluding
all the many-body states which contain a molecule-like bound state
for any two atoms with unlike spins. Strictly speaking, this is a
conjecture which stems from a two-body picture. We have examined this
conjecture using the exact three-fermion energy spectrum near the
resonance. We have found some horizontal energy levels that may be
identified as the energy spectrum of three {}``repulsively'' interacting
fermions, as well as some vertical energy levels involving a tightly-bound
molecule. However, many avoided crossings between horizontal and vertical
levels make it difficult to unambiguously identify the energy spectrum
of a repulsive Fermi system.

For three {}``repulsively'' interacting fermions in a harmonic trap,
we have shown that close to the resonance, the ground state energy
is higher than that of three fully polarized fermions. This is an
indication of the existence of itinerant ferromagnetism in a trapped
strongly repulsively interacting Fermi gas. We have also considered
the possibility of itinerant ferromagnetism at high temperatures.
We have found that it does not exist in the regime where a quantum
virial expansion is applicable. This gives an upper bound ($\sim T_{F}$)
for the critical ferromagnetic transition temperature.

Our high-temperature equations of state of a strongly repulsively
interacting Fermi gas have a number of potential applications. We
anticipate that these results can provide an unbiased benchmark for
future quantum Monte Carlo simulations of strongly repulsively interacting
Fermi gases at high temperatures \cite{akkineni,bulgac,burovski},
using either hard-sphere interatomic potentials or resonance interactions.
These results are also directly testable in future experimental measurements,
as inspired by the most recent thermodynamics measurement at ENS that
have already confirmed our predicted second and third virial coefficients
for strongly attractively interacting fermions \cite{ensexpt}. Our
exact three-fermion solutions in 3D harmonic traps will also be useful
for understanding the dynamical properties of strongly interacting
Fermi gases at high temperatures, by applying a similar quantum virial
expansion for the dynamic structure factors \cite{dsf} and single-particle
spectral functions \cite{akw}.

These exact solutions of three interacting particles can be generalized
to other dimensions, by adopting a suitable Bethe-Peierls boundary
condition for the contact interactions. Of particular interest is
the case of two dimensions, where the reduction of the spatial dimensionality
increases the role of fluctuations and therefore imposes severe challenges
for theoretical studies. The three-body solutions in 2D and the resulting
high-temperature equations of state of strongly interacting systems
will be given elsewhere, and provide a useful starting point to understanding
more sophisticated collective phenomena such as the Berezinsky-Kosterlitz-Thouless
transition and non-Fermi-liquid behavior.

\textit{Note added}: On finishing this manuscript, we are aware of
a very recent work by Daily and Blume \cite{blume2010}, in which
the energy spectrum of three and four fermions has been calculated
using hyperspherical coordinates with a stochastic variational appoach.
Our exact results are in excellent agreement with theirs when there
is an overlap.
\begin{acknowledgments}
This work was supported in part by the ARC Centre of Excellence, ARC
Discovery Project No. DP0984522 and No. DP0984637, NSFC Grant No.
10774190, and NFRPC (Chinese 973) Grant No. 2006CB921404 and No. 2006CB921306. 
\end{acknowledgments}
\appendix

\section{Calculation of $C_{nn^{\prime}}$}

In this appendix, we outline the details of how to construct the matrix
element $C_{nn^{\prime}}$ in Eq. (\ref{amat}), which is given by,
\begin{equation}
C_{nn^{\prime}}\equiv\int\limits _{0}^{\infty}\rho^{2}d\rho R_{nl}\left(\rho\right)R_{n^{\prime}l}\left(\frac{\rho}{2}\right)\psi_{2b}^{rel}(\frac{\sqrt{3}}{2}\rho;\nu_{l,n^{\prime}}),\end{equation}
 where \begin{equation}
R_{nl}\left(\rho\right)=\sqrt{\frac{2n!}{\Gamma\left(n+l+3/2\right)}}\rho^{l}e^{-\rho^{2}/2}L_{n}^{\left(l+1/2\right)}\left(\rho^{2}\right),\end{equation}
 is the radial wave function of an isotropic 3D harmonic oscillator
and the two-body relative wave function is \begin{equation}
\psi_{2b}^{rel}=\Gamma(-\nu_{l,n^{\prime}})U(-\nu_{l,n^{\prime}},\frac{3}{2},\frac{3}{4}\rho^{2})\exp(-\frac{3}{8}\rho^{2}).\end{equation}
 Here, for convenience we have set $d=1$ as the unit of length. $L_{n}^{\left(l+1/2\right)}$
is the generalized Laguerre polynomial and $U$ is the second Kummer
confluent hypergeometric function. A direct integration for $C_{nn^{\prime}}$
is difficult, since the second Kummer function has a singularity at
the origin. The need to integrate for different values of $\nu_{l,n^{\prime}}$
also causes additional complications.

It turns out that a better strategy for the numerical calculations
is to write, \begin{equation}
\psi_{2b}^{rel}=\sum_{k=0}^{\infty}\frac{1}{k-\nu_{l,n^{\prime}}}\sqrt{\frac{\Gamma\left(k+3/2\right)}{2k!}}R_{k0}\left(\frac{\sqrt{3}}{2}\rho\right),\end{equation}
 by using the exact identity, \begin{equation}
\Gamma(-\nu)U(-\nu,\frac{3}{2},x^{2})=\sum_{k=0}^{\infty}\frac{L_{k}^{1/2}\left(x^{2}\right)}{k-\nu}.\end{equation}
 Therefore, we find that \begin{equation}
C_{nn^{\prime}}=\sum_{k=0}^{\infty}\frac{1}{k-\nu_{l,n^{\prime}}}\sqrt{\frac{\Gamma\left(k+3/2\right)}{2k!}}C_{nn^{\prime}k}^{l},\end{equation}
 where \begin{equation}
C_{nn^{\prime}k}^{l}\equiv\int\limits _{0}^{\infty}\rho^{2}d\rho R_{nl}\left(\rho\right)R_{n^{\prime}l}\left(\frac{\rho}{2}\right)R_{k0}\left(\frac{\sqrt{3}}{2}\rho\right)\end{equation}
 can be calculated to high accuracy with an appropriate integration
algorithm. In checking convergence of the summation over $k$, we
find numerically that for a cut-off $n_{\max}$ (i.e., $n,n^{\prime}<n_{\max}$),
$C_{nn^{\prime}k}^{l}$ vanishes for a sufficient large $k>k_{\max}\sim4n_{\max}$.

In practical calculations, we tabulate $C_{nn^{\prime}k}^{l}$ for
a given total relative angular momentum. The calculation of $C_{nn^{\prime}}$
for different values of $\nu_{l,n^{\prime}}$ then reduces to a simple
summation over $k$, which is very efficient. Numerically, we have
confirmed that the matrix $C_{nn^{\prime}}$ is symmetric, i.e., $C_{nn^{\prime}}=C_{n^{\prime}n}$.

\section{Calculation of $s_{l,n}$}

The calculation of $s_{l,n}$ seems straightforward by using the Bethe-Peierls
boundary condition in hyperspherical coordinates (\ref{BP3eHyper}).
However, we find that numerical accuracy is low for large $n$ and
$l$ due to the difficulty of calculating the hypergeometric function
$_{2}F_{1}$ accurately using IEEE standard precision arithmetic.
We have therefore utilized MATHEMATICA software that can perform analytical
calculations with unlimited accuracy. For this purpose, we introduce
$\Delta s_{l,n}=s_{l,n}-\bar{s}_{l,n}$. After some algebra, we find
the following boundary condition for $t\equiv\Delta s_{l,n}/2$, \begin{equation}
\sin\left(\pi t\right)=\sqrt{\frac{\pi}{3}}\frac{\left(-1\right)^{n+l}\Gamma\left(n+l+1+t\right)}{2^{l}\Gamma\left(l+\frac{3}{2}\right)\Gamma\left(n+1+t\right)}f\left(t\right),\label{dsln}\end{equation}
 where we have defined a function \begin{equation}
f\left(t\right)\equiv\text{ }_{2}F_{1}\left(-n-t,n+l+1+t,l+\frac{3}{2};\frac{1}{4}\right).\end{equation}
 The above equation can be solved using the MATHEMATICA routine {}``FindRoot'',
by seeking a solution around $t=0$. It is also easy to write a short
program to solve Eq. (\ref{dsln}) continuously for $n<n_{\max}=512$
and $l<l_{\max}=512$. In a typical current PC, this takes several
days. The results can be tabulated and stored in a file for further
use.


\begin{thebibliography}{38}
\bibitem{braaten} E. Braaten and H. Hammer, Phys. Rep. \textbf{428},
259 (2006).

\bibitem{bloch} I. Bloch, J. Dalibard, and W. Zwerger, Rev. Mod.
Phys. \textbf{80}, 885 (2008).

\bibitem{giorgini} S. Giorgini, L. P. Pitaevskii, S. Stringari, Rev.
Mod. Phys. \textbf{80}, 1215 (2008).

\bibitem{unitaritycmp} H. Hu, X.-J. Liu, and P. D. Drummond, arXiv:
1001.2085; to be published in New J. Phys. (2010).

\bibitem{chin} C. Chin, R. Grimm, P. Julienne, and E. Tiesinga, Rev.
Mod. Phys. \textbf{82}, 1225 (2010).

\bibitem{liu2009} X.-J. Liu, H. Hu, and P. D. Drummond, Phys. Rev.
Lett. \textbf{102}, 160401 (2009).

\bibitem{ensexpt} S. Nascimbène, N. Navon, K. Jiang, F. Chevy, and
C. Salomon, Nature \textbf{463}, 1057 (2010).

\bibitem{mitexpt} G.-B. Jo, Y.-R. Lee, J.-H. Choi, C. A. Christensen,
T. H. Kim, J. H. Thywissen, D. E. Pritchard, and W. Ketterle, Science
\textbf{325}, 1521 (2009).

\bibitem{blume2002} D. Blume and C. H. Greene, Phys. Rev. A \textbf{66},
013601 (2002).

\bibitem{stecher2007} J. von Stecher and C. H. Greene, Phys. Rev.
Lett. \textbf{99}, 090402 (2007).

\bibitem{blume2009} D. Blume and K. M. Daily, Phys. Rev. A \textbf{80},
053626 (2009).

\bibitem{petrov} D. S. Petrov, C. Salomon, and G. V. Shlyapnikov,
Phys. Rev. Lett. \textbf{93}, 090404 (2004).

\bibitem{physics} For a brief review, see for example, F. Ferlaino
and R. Grimm, Physics \textbf{3}, 9 (2010).

\bibitem{stoner1938} E. C. Stoner, Proc. R. Soc. London. Ser. A \textbf{165},
372 (1938).

\bibitem{macdonald2005} R. A. Duine and A. H. MacDonald, Phys. Rev.
Lett. \textbf{95}, 230403 (2005).

\bibitem{zhang2008} S. Zhang, H.-H. Hung, C. Wu, arXiv:0805.3031
(2008).

\bibitem{leblanc2009} J. L. LeBlanc, J. H. Thywissen, A. A. Burkov,
and A. Paramekanti, Phys. Rev. A \textbf{80}, 013607 (2009).

\bibitem{conduit2009a} G. J. Conduit and B. D. Simons, Phys. Rev.
Lett. \textbf{103}, 200403 (2009).

\bibitem{conduit2009b} G. J. Conduit, A. G. Green, and B. D. Simons,
Phys. Rev. Lett. \textbf{103}, 207201 (2009).

\bibitem{zhai2009} H. Zhai, Phys. Rev. A \textbf{80}, 051605(R) (2009).

\bibitem{cui2010} X. Cui and H. Zhai, Phys. Rev. A \textbf{81}, 041602(R)
(2010).

\bibitem{spilati} S. Pilati, G. Bertaina, S. Giorgini, and M. Troyer,
arXiv: 1004.1169 (2010).

\bibitem{chang} S.-Y. Chang, M. Randeria, and N. Trivedi, arXiv:
1004.2680 (2010).

\bibitem{dong2010} H. Dong, H. Hu, X.-J. Liu, and P. D. Drummond,
arXiv: 1004.5443 (2010).

\bibitem{ho2004a} T.-L. Ho, Phys. Rev. Lett. \textbf{92}, 090402
(2004).

\bibitem{natphys} H. Hu, P. D. Drummond, and X.-J. Liu, Nature Phys.
\textbf{3}, 469 (2007).

\bibitem{busch} T. Busch, B. G. Englert, K. Rzazewski, and M. Wilkens,
Found. Phys. \textbf{28}, 549 (1998).

\bibitem{wernerprl} F. Werner and Y. Castin, Phys. Rev. Lett. \textbf{97},
150401 (2006).

\bibitem{wernerpra} F. Werner and Y. Castin, Phys. Rev. A \textbf{74},
053604 (2006).

\bibitem{footnote} F. Werner, PhD thesis, École Normale Supérieure
(2008).

\bibitem{kestner} J. P. Kestner and L.-M. Duan, Phys. Rev. A \textbf{76},
033611 (2007).

\bibitem{ho2004b} T.-L. Ho and E. J. Mueller, Phys. Rev. Lett. \textbf{92},
160404 (2004).

\bibitem{akkineni} V. K. Akkineni, D. M. Ceperley, and N. Trivedi,
Phys. Rev. B \textbf{76}, 165116 (2007).

\bibitem{bulgac} A. Bulgac, J. E. Drut, and P. Magierski, Phys. Rev.
Lett. \textbf{96}, 090404 (2006).

\bibitem{burovski} E. Burovski, N. Prokof'ev, B. Svistunov, and M.
Troyer, Phys. Rev. Lett. \textbf{96}, 160402 (2006).

\bibitem{dsf}H. Hu, X.-J. Liu, and P. D. Drummond, Phys. Rev. A \textbf{81},
033630 (2010).

\bibitem{akw}H. Hu, X.-J. Liu, and P. D. Drummond, Phys. Rev. Lett.
\textbf{104}, 240407 (2010).

\bibitem{blume2010}K. M. Daily and D. Blume, Phys. Rev. A \textbf{81},
053615 (2010); also avaialbe as arXiv: 1006.0769 (2010).
\end{thebibliography}
\end{document}